\documentclass{jpsj3}
\usepackage{txfonts}
\usepackage{graphicx}% Include figure files
\usepackage{amsmath}
\usepackage{braket}
\usepackage{here}
\usepackage{siunitx}
\usepackage{csquotes}

\title{Evaluation of quantum entanglement state\\ between photoelectron spin and emitted photon polarization\\ in spin and polarization resolved XEPECS of $\rm Ti_{2}O_{3}$}

\author{Ryo B. Tanaka\thanks{su23179r@st.omu.ac.jp}$^{1}$, Goro Oohata$^{2,3,4}$ and Takayuki Uozumi$^{1}$}
\inst{$^{1}$Department of Physics and Electronics, Graduate School of Engineering, \\Osaka Metropolitan University,\\ Sakai, Osaka 599-8531, Japan \\
$^{2}$Department of Physics, Graduate School of Science,\\ Osaka Metropolitan University,\\ Sakai, Osaka 599-8531, Japan \\
$^{3}$Department of Physics Nambu Yoichiro Institute of Theoretical\\ and Experimental Physics (NITEP),\\ Osaka Metropolitan University,\\ Sugimoto 3-3-138, Sumiyoshi-ku, Osaka 558-8585, Japan \\
$^{4}$Research Institute for Light-induced Acceleration System (RILACS),\\ Osaka Metropolitan University,\\ 1-2 Gakuencho, Nakaku, Sakai, Osaka, 599-8570, Japan}

\abst{We theoretically investigated the mechanism of quantum entanglement between the spin of photoelectrons and linear polarization of emitted X-ray photons in the 3$d\rightarrow\ $2$p$ XEPECS process for $\rm Ti_{2}O_{3}$.  In the calculation, we used a realistic $\rm TiO_{6}$-type cluster model with the full multiplet structure of the Ti ion and the charge-transfer effect between the Ti 3$d$ and ligand O 2$p$ orbitals. We found that quantum entanglement occurs between the spin of photoelectrons and linear polarization of emitted X-ray photons and that it depends on the angular geometry in the XEPECS process.
In addition, we found that the degree of spin and polarization entanglement decreases as the Ti 3$d\ $- O 2$p$ hybridization becomes stronger and as the crystal field modifies the electronic states in terms of the tangle, an index for the degree of entanglement. These results highlight the crucial role of the charge transfer and crystal field effects in determining entanglement properties in real material systems.}

\begin{document} 
\maketitle

\section{Introduction}

Quantum entanglement, which expresses spooky nonlocal correlations, is a necessary resource in quantum information and communication technology\cite{HORODECKI}. Recently, electron-photon entanglement has gained importance for developing quantum teleportation and quantum repeater technology\cite{TOGAN,KOSAKA,TSURUMOTO,SEKIGUCHI,REISERER}. Actually, electron-photon entanglement has been studied in the low-energy region\cite{TOGAN,KOSAKA,TSURUMOTO,SEKIGUCHI}. However, there have been a few reports in the high-energy region\cite{CHOWDHURY,RTANAKA}. 

Recently, several experiments have been conducted related to quantum entanglement in the X-ray wavelength region owing to the development of synchrotron radiation systems\cite{SHWARTZ1, SHWARTZ2, RIVERA, SOFER, STRIZHEVSKY, ZHANG, HARTLEY}. Most of the experiments are on the production of X-ray photon pairs, and there are a few studies on photoelectron-emitted X-ray photon pairs. 
An example of such studies is the recent reports considering the entanglement between the electrons and photons using high-energy Compton scattering\cite{CHOWDHURY} and electron microscopy\cite{YANAGIMOTO}.

In this study, we treat the XEPECS process, i.e., XES (X-ray emission
spectroscopy) $\&$ cXPS (core-level X-ray photoemission spectroscopy) coincidence spectroscopy\cite{STANAKA}, one of the X-ray spectroscopic processes, as a method to produce a pair of electron and photon. XEPECS simultaneously measures the kinetic energy of photoelectrons and the energy of X-ray photons emitted from the same atom. 
In this paper, to investigate the quantum entanglement between the spin of photoelectrons and polarization of emitted X-ray photons, we treat photoelectron spin- and emitted X-ray photon polarization-resolved XEPECS. In addition, to investigate the geometric effects of the quantum entanglement states, we also resolve the incident angle and polarization of the incident photons, as well as the emission angle of the photoelectrons and emitted X-ray photons.

Using a simple atomic model, so far, we theoretically investigated the quantum entanglement between the spin of photoelectron and linear polarization of emitted X-ray photon by 2$p\rightarrow\ $1$s$ XEPECS process\cite{RTANAKA}. As a result, we found that quantum entanglement between the spin of the photoelectron and linear polarization of the emitted X-ray photon is generated. 
Furthermore, the degree of entanglement has a strong emission angle dependence on X-ray photons.
In spite of a simple model, we showed that quantum entanglement emerges between the photoelectrons and emitted X-ray photons in the XEPECS process.
However, the simple atomic model does not include both the charge-transfer effect due to the core-excitation
and the full-multiplet interactions, which can strongly affect the spectral structure and polarization dependence\cite{OKADA, UOZUMI, MATSUBARA1}.
Thus, a theoretical investigation using a realistic model is strongly desired.

To address this, we choose $\rm Ti_{2}O_{3}$ as a target system in this paper, which is a typical transition metal oxide of the $3d^1$ system and is a direct extension of our previous $sp$-model\cite{RTANAKA}, and use a charge-transfer $\rm TiO_{6}$-type cluster model\cite{OKADA, UOZUMI, MATSUBARA1} considering the full-multiplet coupling effect.
The purpose of this study is to clarify the effect of charge transfer on the degree of quantum entanglement between the photoelectron spin and emitted X-ray photon polarization in XEPECS. 
In addition, we investigate the emission angle dependence of the X-ray photons on spin and polarization entanglement and the role of the crystal field in modifying electronic states.
Needless to say, this study focuses on the processes of photoelectron emission and X-ray emission occurring in the same atom. 
Actually, coincidence measurements have been conducted in recent experiments\cite{JANNIS1, JANNIS2} for the combined spectroscopy between the electron energy loss and energy dispersive X-ray. As for XEPECS experiments, however, no experiments have been performed, probably due to technical difficulties. Thus, the present study is positioned as a theoretical prediction.

\section{Theoritical method}
\begin{figure}[H]
    \centering
    \includegraphics[scale=0.45]{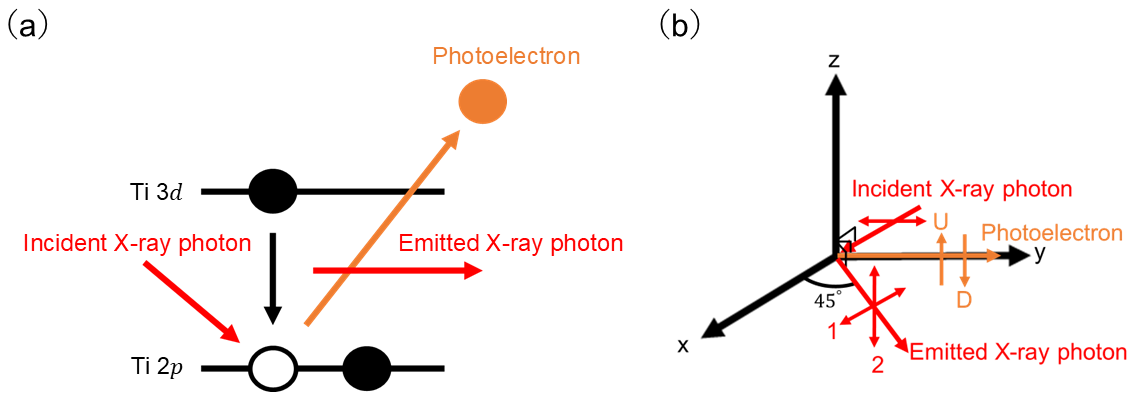}
    \caption{(a) A schematic view of the photo-induced transition between the 2$p$ and 3$d$ levels with emissions of photoelectron and emitted X-ray photon. (b) The geometrical setting of the linearly polarized incident photon, the photoelectron with spin (U for up and D for down), and the emitted photon with polarization (1 for $\lambda_{1}$ and 2 for $\lambda_{2}$).}
    \label{Fig.1}
\end{figure}

In this study, we focus on the quantum entanglement between the spin of photoelectrons and polarization of emitted X-ray photons in XEPECS of $\rm Ti_{2}O_{3}$.
Here we consider the XEPECS process shown in Fig. \ref{Fig.1}(a), where a 2$p$ core electron is emitted as a photoelectron by an incident X-ray and an X-ray photon is emitted through the Ti 3$d\rightarrow\ $2$p$ radiative decay of the 2$p$ core hole. 
In the calculation, for the target compound $\rm Ti_{2}O_{3}$, we adopt a $\rm TiO_{6}$-type cluster model with the $O_{h}$ symmetry, including the full-multiplet structure of the Ti ion and the Ti 3$d$ - O 2$p$ charge transfer.

To describe the participating electrons in the process, we consider the Hamiltonian,
\begin{align}
\label{ham_0}
    H_{0} = H_{CL} + H_{PE},
\end{align}
where $H_{CL}$ is for the $\rm TiO_{6}$ cluster and $H_{PE}$ the photoelectrons.
$H_{CL}$ describes the Ti 3$d$, O 2$p$ and core states, and is given by
\begin{align}
\label{ham_clu}
    H_{CL} = &\sum_{\gamma}\epsilon_{d}(\gamma)d_{\gamma}^{\dag}d_{\gamma} + \epsilon_{P}\sum_{\gamma}P_{\gamma}^{\dag}P_{\gamma} + \sum_{\gamma}V(\gamma)(d_{\gamma}^{\dag}P_{\gamma} + P_{\gamma}^{\dag}d_{\gamma})\notag\\
    &+ U_{dd}\sum_{\gamma>\gamma'}n_{d_{\gamma}}n_{d_{\gamma'}} - U_{dc}\sum_{\gamma,\xi}n_{d_{\gamma}}(1-c_{\xi}^{\dag}c_{\xi}) + H_{mult},
\end{align}
where $d_{\gamma}^{\dag}$, $c_{\xi}^{\dag}$ and $P_{\gamma}^{\dag}$ are the electron creation operators for the Ti 3$d$, Ti 2$p$ core orbitals and O 2$p$ molecular orbitals, respectively, $n_{d_{\gamma}}$ is the number operator for $d_{\gamma}$ state, and $\gamma$ and $\xi$ mean the combined indices between the orbital and spin states.
The first and second terms in Eq. (\ref{ham_clu}) represent the one-body energies for the Ti 3$d$ and O 2$p$ states, respectively. The third, fourth and fifth terms describe the hybridization between the Ti 3$d$ and O 2$p$ molecular orbitals, the Coulomb interaction between the 3$d$ electrons and core-hole potential between the 2$p$ core hole and 3$d$ electron, respectively. 
$H_{mult}$ consists of the spin-orbit interactions for 3$d$ and 2$p$ core states and the Coulombic multipole parts of the 3$d$ - 3$d$ and the 3$d$ - 2$p$ interaction and describes the full-multiplet interaction in the Ti ion.

The cluster model includes the solid-state parameters, the charge transfer energy $\Delta(\equiv\epsilon_{d}(\gamma)-\epsilon_{P})$, the Ti 3$d$ crystal-field splitting $10Dq(=\epsilon_{d}(e_{g})-\epsilon_{d}(t_{2g}))$, $V(\gamma)$, $U_{dd}$ and $U_{dc}$ as adjustable parameters. Actually, we adopt here the parameter values $\Delta=6.5\ \rm eV$, $10Dq=0.5\ \rm eV$, $V(e_{g})=3.0\ \rm eV$ ($=-2V(t_{2g})$), $U_{dd}=4.5\ \rm eV$ and $U_{dc}=5.3\ \rm eV$  which are obtain in the analysis of Ti 2$p$ XPS for $\rm Ti_{2}O_{3}$\cite{UOZUMI}.
Similarly, we use the parameter values listed in ref. 19 for the multiplet interactions in $H_{mult}$, which are estimated through the Hatree-Fock-Slater calculation for the Ti ion.

The second term in Eq. (\ref{ham_0}) is represented as 
\begin{align}
\label{ham_PE}
    H_{PE} = \sum_{\vec{k},m,\sigma}\varepsilon_{\vec{k}} b_{\vec{k}m\sigma}^{\dag}b_{\vec{k}m\sigma},
\end{align}
where $\varepsilon_{\vec{k}}$ is the kinetic energy of the photoelectron with the wave-number vector $\vec{k}$ and $b_{\vec{k}m\sigma}^{\dag}$ is the corresponding creation operator for the photoelectron with the orbital magnetic quantum number $m$ and the spin $\sigma$. 
As in the previous paper\cite{RTANAKA}, 
$m$ and up/down for the photoelectron spin state is referred to $z$ axis as shown in Fig. \ref{Fig.1}(b).
In the present study for the angle-resolved XEPECS, however, we consider only the photoelectrons emitted to a specified direction, and thus $\vec{k}$ is dropped in the following discussion for simplicity.

The quantum states of the system described by the unperturbed Hamiltonian in Eq. (\ref{ham_0}) undergo a perturbative transition due to the electron-photon interactions,
\begin{align}
\label{V}
    V = \int d\varepsilon\ M_{PE}e^{-i\Omega t} + M_{ph}e^{i\omega t} + \rm h.c.
\end{align}
In Eq. (\ref{V}), the first and second terms represent the emission of the photoelectrons from the 2$p$ core by incident X-ray irradiations and
the emission of X-ray photons due to the Ti 3$d\rightarrow\ $2$p$ radiative decay of the 2$p$ core hole, respectively, and $M_{PE}$ and $M_{ph}$ are described by 
\begin{align}
\label{M_PE}
    &M_{PE}=\sum_{m,m',\sigma}Y_{Dm}(\theta_{\rm e},\phi_{\rm e})\braket{D_{m}|\vec{e}_{\lambda_{\rm in}}\cdot\vec{r}|p_{m'}}b_{m\sigma}^{\dag}p_{m'\sigma}a_{\lambda_{\rm in}},\\
    \label{M_ph}&M_{ph}=\sum_{\lambda,m,m',\sigma}\braket{p_{m}|\vec{e}_{\lambda}\cdot\vec{r}|d_{m'}}p_{m\sigma}^{\dag}d_{m'\sigma}a_{\lambda}^{\dag}.
\end{align}
In Eq. (\ref{M_PE}), $Y_{Dm}(\theta_{\rm e},\phi_{\rm e})$ is the spherical harmonic function with the angular momentum $l=D$ and the orbital magnetic quantum number $m$.
It should be noted that we consider here only the dipole component in transition amplitude $\braket{D_{m}|\vec{e}_{\lambda_{\rm in}}\cdot\vec{r}|p_{m'}}$ with the $D$-symmetry component of photoelectron, and the spherical harmonics as a function of the emission angle $(\theta_{\rm e}, \phi_{\rm e})$ of the photoelectron is attached.
In addition, $a_{\lambda_{\rm in}}$ is the annihilation operator of an incident photon with energy $\Omega$ and polarization vector $\vec{e}_{\lambda_{\rm in}}$, where $\lambda_{\rm in}$ denotes the polarization mode under the Coulomb gauge condition. 
In the 3$d\rightarrow\ $2$p$ dipole transition operator in Eq.(\ref{M_ph}), $a_{\lambda}^{\dag}$ is the creation operator for the emitted photon with the energy $\omega$ and the polarization vector $\vec{e}_{\lambda}$, where $\lambda$ means the polarization mode of the emitted photon under the Coulomb gauge condition.
Actually, we consider the geometry as illustrated in Fig. \ref{Fig.1}(b).
For the incident photon, we set the incidence angles as $(\theta_{\rm in}, \phi_{\rm in}) = (90^{\circ}, 0^{\circ})$ and consider linear polarization $\lambda_{\rm in}$ as parallel to $y$ axis.
Then, for the photoelectron, we set the emission angles as $(\theta_{\rm e}, \phi_{\rm e}) = (90^{\circ}, 90^{\circ})$. 
As for the emitted photon, we set the emission angles as $(\theta_{\rm out}, \phi_{\rm out}) = (90^{\circ}, 45^{\circ})$ and consider the linear polarization as $\lambda = 1$ and $2$.
Thus, we consider the situation where an incident photon goes along the $x$ axis, the photoelectron is emitted along the $y$ axis, and the emitted photon goes to the direction $\phi_{\rm out}=45^{\circ}$ in the $xy$ plane.

The initial state $\ket{g}$ of the system is described by the product of the subsystems $\ket{g}_{\rm clu}$ and $\ket{\lambda_{\rm in}}_{\rm ph}$ as
\begin{align}
    \label{g_clu} 
    \ket{g}=\ket{g}_{\rm clu}\ket{\lambda_{\rm in}}_{\rm ph}
    =(\alpha_{0}\ket{d^{1}}+\alpha_{1}\ket{d^{2}\underline{L}}+\alpha_{2}\ket{d^{3}\underline{L}^{2}})\ket{\lambda_{\rm in}}_{\rm ph},
\end{align}
where $\ket{g}_{\rm clu}$ represents the ground state of the $\rm TiO_{6}$ cluster and $\ket{\lambda_{\rm in}}_{\rm ph}$ the incident photon with polarization $\lambda_{\rm in}$, respectively.
As in ref. 19, $\ket{g}_{\rm clu}$ is described as a linear combination of three basis configurations $d^{1}$, $d^{2}\underline{L}$ and $d^{3}\underline{L}^{2}$, where $\underline{L}$ means a ligand hole, and these configurations mix with each other through the $p$-$d$ hybridization in Eq. (\ref{ham_clu}).

To obtain the density matrix for the combined state between the photoelectron spin and emitted photon polarization, we need to calculate the wave function "just before observation" in the 3$d\rightarrow\ $2$p$ XEPECS process, which is a function of the time $t$, the binding energy $E_{B}$ of the photoelectron and the emitted photon energy $\omega$.
In the rotating-wave approximation\cite{LOUDON}, the wave function is obtained within the framework of the coherent second-order optical process and is expressed as
\begin{align}
\label{wave func}
    \ket{\psi(t,E_{B},\omega)}=-i\sum_{f,i,\sigma,\lambda}\frac{\braket{f_{\sigma\lambda}|M_{ph}|i_{\sigma}}\braket{i_{\sigma}|M_{PE}|g}}{E_{B}+E_{g}-E_{i_{\sigma}}+i\Gamma_{2p}}\ket{f_{\sigma\lambda}}\int_{-\infty}^{t}dt'e^{i(E_{f_{\sigma\lambda}}+\omega-E_{B}-E_{g})t'}.
\end{align}
Here $\ket{g}$, $\ket{i_{\sigma}}$ and $\ket{f_{\sigma\lambda}}$ denote the initial, intermediate and final states with the energies $E_{g}$, $E_{i_{\sigma}}$ and $E_{f_{\sigma\lambda}}$, respectively, and $\Gamma_{2p}$ is the inverse lifetime of the 2$p$ core hole. Note that $\ket{i_{\sigma}}$ and $\ket{f_{\sigma\lambda}}$ are the states with the photoelectron with spin $\sigma$ and the emitted photon with the polarization $\lambda$. In the present study, we treat the wave function as a stationary state by taking the limit $t \rightarrow \infty$ for simplicity. Thus, we omit $t$ in the notation $\ket{\psi}$ below.
In order to obtain the density matrix $\rho_{\sigma\lambda,\sigma'\lambda'}$ for the photoelectron spin state and emitted photon polarization state, we start from the projector,
\begin{align}
\label{chi}    \ket{\psi(E_{B},\omega)}\bra{\psi(E_{B},\omega)}    =\sum_{f,f',\sigma,\sigma',\lambda,\lambda'}\braket{f_{\sigma\lambda}|M_{ph}|x_{\sigma}(E_{B})}\braket{x'_{\sigma'}(E_{B})|M_{ph}^{\dag}|f'_{\sigma'\lambda'}}\ket{f_{\sigma\lambda}}\bra{f'_{\sigma'\lambda'}}\notag\\
    \hspace{3.5cm}\times\delta(E_{B}+E_{g}-\omega-E_{f_{\sigma\lambda}})\delta(E_{B}+E_{g}-\omega-E_{f'_{\sigma'\lambda'}}),
\end{align}
where $\ket{x_{\sigma}(E_{B})}$ is
\begin{align}
\label{x}
    \ket{x_{\sigma}(E_{B})}\equiv\sum_{i}\frac{\braket{i_{\sigma}|M_{PE}|g}}{E_{B}+E_{g}-E_{i_{\sigma}}+i\Gamma_{2p}}\ket{i_{\sigma}}.
\end{align}
From the two delta functions in Eq. (\ref{chi}), we obtain the relationship $E_{f_{\sigma\lambda}}=E_{f'_{\sigma'\lambda'}}$.
Next, we treat the final state $\ket{f_{\sigma\lambda}}$ as the direct product of subsystems $\ket{f}_{\rm clu}$, $\ket{\sigma}_{\rm el}$ and $\ket{\lambda}_{\rm ph}$ for the cluster final states, the photoelectron spin and the emitted photon, respectively. By taking the trace sum over $\ket{f}_{\rm clu}$ of Eq. (\ref{chi}), we obtain the expression for the density matrix
\begin{align}
    \label{rho}
    \rho_{\sigma\lambda,\sigma'\lambda'}(E_{B},\omega)=\frac{1}{\rm Tr(\rho_{\sigma\lambda,\sigma\lambda})}\sum_{\sigma,\sigma',\lambda,\lambda'}\braket{f_{\sigma\lambda}|M_{ph}|x_{\sigma}(E_{B})}\braket{x'_{\sigma'}(E_{B})|M_{ph}^{\dag}|f_{\sigma'\lambda'}}\notag\\
    \times\delta(E_{B}+E_{g}-\omega-E_{f_{\sigma\lambda}})
    \ket{\sigma}_{\rm el}\ket{\lambda}_{\rm ph}\bra{\sigma'}_{\rm el}\bra{\lambda'}_{\rm ph}.
\end{align}
Actually, we calculate numerically $\rho_{\sigma\lambda,\sigma'\lambda'}$ by setting $E_{B}$ and $\omega$.

To evaluate the quantum entanglement state between the spin of the photoelectron and the linear polarization of the emitted photon, we use the indices of the quantum entanglement, fidelity, tangle and linear entropy in the present study.
The fidelity $F$ is defined as 
\begin{align}
\label{F}
    F=\braket{\psi|\rho|\psi}
\end{align}
using the density matrix $\rho$ and the so-called ideal state $\ket{\psi}$.\cite{EDAMATSU,OOHATA}
The tangle $T$, which quantifies the degree of entanglement, is defined using the concurrence $C$ as $T=C^{2}$, where $C$ is defined as
\begin{align}
\label{conc}
    C=\rm{max}(\sqrt{\Lambda_{1}}-\sqrt{\Lambda_{2}}-\sqrt{\Lambda_{3}}-\sqrt{\Lambda_{4}},0).
\end{align}
Here, $\Lambda_{i}\rm s\ (i=1\sim4)$ are the eigenvalues arranged in decreasing order of the matrix
\begin{align}
    \rho_{\sigma\lambda,\sigma'\lambda'}(\sigma_{y}\otimes\sigma_{y})\rho_{\sigma\lambda,\sigma'\lambda'}^{*}(\sigma_{y}\otimes\sigma_{y}).
\end{align}
The linear entropy $S_{L}$\cite{WOOTTERS,COFFMAN,JAMES,MUNRO,MUNRO2}, which quantifies the degree of mixture of a quantum state, is given by 
\begin{align}
    S_{L}=\frac{4}{3}(1-\rm{Tr}(\rho^{2})).
\end{align}

\section{Results and discussion}
We discuss the quantum entanglement between the spin of photoelectrons and polarization of emitted photons in 3$d\rightarrow\ $2$p\ $XEPECS of $\rm Ti_{2}O_{3}$, using numerical results for the density matrix, fidelity, tangle and linear entropy.
The values of the parameters in the cluster model are the same as those used in ref. 19 and are listed in Table \ref{table}.
Here $\Delta$ is the charge transfer energy, $U_{dd}$ is the Coulomb interaction between 3$d$ electrons, $V(e_{g})$ is the $p$-$d$ hybridization for the $e_{g}$ orbitals, $U_{dc}(2p)$ is the 2$p$ core-hole potential for 3$d$ electrons and  $10Dq$ is the crystal field splitting energy of $3d$ levels.
An empirical relation $2V(t_{2g}) = -V(e_{g})$ is also assumed for the $p$-$d$ hybridization for the $t_{2g}$ orbitals \cite{UOZUMI}.

Under a crystal field with $O_{h}$ symmetry, the basis system for Hamiltonian in Eq.(\ref{ham_clu}) divides into four subspaces.
For the basis system with the magnetic quantum number $M$, each basis systems with $\cdots, M-4, M, M+4,\cdots$ constitute a subspace, respectively, because of the crystal-field operator $C^{4}_{0} + \sqrt{5/14}(C^{4}_{4}+C^{4}_{-4})$ for $3d$ electrons.
These four subspaces are distinguished by the magnetic quantum number $M = -3/2, -1/2, 1/2$ and $3/2$ as labels, respectively.
In this case, the minimized energy state with each of the four subspaces degenerates.
This is due to the fourfold degeneracy of $d^{1}$ system under the spin-orbit interaction and crystal field.
The present study uses the subspace with $M=-1/2$ as the initial state $\ket{g}$ in Eq.(\ref{g_clu}).
This basis system is suitable for describing the microscopic mechanism of quantum entanglement and aligns with the purpose of the present study. 
Thus, we focus on the basis system with $M=-1/2$ to advance the discussion.
Future comprehensive work will focus on the antiferromagnetic effect of the adjacent arrangement of two Ti ions with $M=-1/2$ and $1/2$, respectively.

\begin{table}[htbp]
\centering
\begin{tabular}{ccccccccc}
\hline
$\Delta$ & $\ $ & $U_{dd}$ & $\ $ & $V(e_{g})$ & $\ $ & $U_{dc}(2p)$ & $\ $ & $10Dq$ \\
\hline
6.5 & $\ $ & 4.5 & $\ $ & 3.0 & $\ $ & 5.3 & $\ $ & 0.5 \\
\hline
\end{tabular}
\caption{Values of parameters (eV) for calculation in $\rm Ti_{2}O_{3}$.}
\label{table}
\end{table}

\begin{figure}[H]
    \centering
    \includegraphics[scale=0.5]{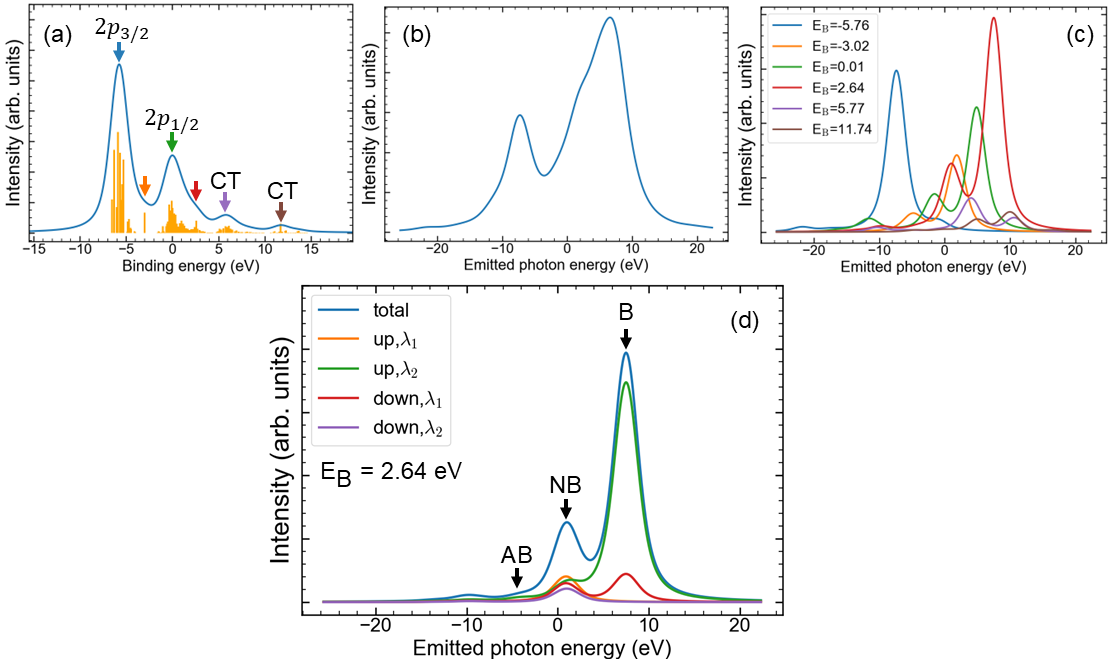}
    \caption{(a) The 2$p$ XPS, (b) the 3$d\rightarrow\ $2$p\ $NXES and (c) the 3$d\rightarrow\ $2$p\ $XEPECS of $\rm Ti_{2}O_{3}$. Each color spectrum in (c) corresponds to the specific binding energy indicated by the arrow with the same color in (a). (d) Photoelectron's spin and emitted photon's polarization-resolved 3$d\rightarrow\ $2$p\ $XEPECS at the unresolved red spectrum in (c).
    The energies of the initial state ($E_{g}$), intermediate state ($E_{i}$), and final state ($E_{f}$) are each defined relative to the configuration averaged energy of the first electronic configuration. Based on these definitions, the horizontal axes are defined according to Eqs. (\ref{XPS}) and (\ref{NXES}): the binding energy is $E_{B} = E_{i} - E_{g}$, and the emitted photon energy is $\omega = E_{B} + E_{g} - E_{f}$.}
    \label{Fig.2}
\end{figure}

First, we set the binding energy $E_{B}(=\Omega-\varepsilon)$, where $\Omega$ is the incident photon energy and $\varepsilon$ is the photoelectron kinetic energy, and the emission energy $\omega$ to calculate the density matrix.
Figs. \ref{Fig.2}(a), 2(b) and 2(c) show the calculated results of 2$p$ XPS, 3$d\rightarrow\ $2$p\ $normal XES (NXES) and 3$d\rightarrow\ $2$p\ $XEPECS for $\rm Ti_{2}O_{3}$, respectively. 
The 2$p$ XPS is calculated using
\begin{align}
\label{XPS}
    F_{\rm XPS}(E_{B})=\sum_{i}\left|\braket{i|M_{PE}|g}\right|^{2}\delta(E_{B}+E_{g}-E_{i}).
\end{align}
The continuous spectrum shown in Fig. \ref{Fig.2}(a) is obtained by convoluting the line spectra from Eq. (\ref{XPS}) with the Lorentzian and Gaussian functions with the half width at half maxima (HWHM) $\Gamma_{\rm L} = 0.7$ eV and $\Gamma_{\rm G} = 0.5$ eV, respectively\cite{UOZUMI}.
In addition, the 3$d\rightarrow\ $2$p\ $NXES and 3$d\rightarrow\ $2$p\ $XEPECS are calculated using 
\begin{align}
\label{NXES}
    F_{\rm NXES}(\omega)=\int dE_{B}\sum_{f}\left|\sum_{i}\frac{\braket{f|M_{ph}|i}\braket{i|M_{PE}|g}}{E_{B}+E_{g}-E_{i}+i\Gamma_{2p}}\right|^{2}\delta(E_{B}+E_{g}-\omega-E_{f}),\\
\label{XEPECS}
    F_{\rm XEPECS}(E_{B},\omega)=\sum_{f}\left|\sum_{i}\frac{\braket{f|M_{ph}|i}\braket{i|M_{PE}|g}}{E_{B}+E_{g}-E_{i}+i\Gamma_{2p}}\right|^{2}\delta(E_{B}+E_{g}-\omega-E_{f}),
\end{align}
where $\Gamma_{2p}$ is the inverse lifetime of the 2$p$ core hole, and here we set $\Gamma_{2p} = 0.5$ eV.
The continuous spectra in Figs. \ref{Fig.2}(b) and (c) are obtained by convoluting Eqs. (\ref{NXES}) and (\ref{XEPECS}) with the Lorenzian ($\Gamma_{\rm L} = 1.0$ eV) and Gaussian ($\Gamma_{\rm G} = 1.0$ eV) functions. These values of $\Gamma_{\rm L}$ and $\Gamma_{\rm G}$ are provisional parameters because there are currently no experimental data for the 3$d\rightarrow\ $2$p\ $XEPECS.

As discussed in ref. 19, the 2$p$ XPS in Fig. \ref{Fig.2}(a) consists of the 2$p$ spin-orbit doublets $2p_{3/2}$ and $2p_{1/2}$ and of their charge-transfer (CT) satellites, and is plotted against the relative binding energy $E_{B}$. 
In Fig. \ref{Fig.2}(b), the 3$d\rightarrow\ $2$p\ $NXES shows the two-peak structure with the broad main peak on the high-energy side. 
In the calculation of the 3$d\rightarrow\ $2$p\ $NXES in Fig. \ref{Fig.2}(b), we used binding energies in increments of 0.2 eV over the $E_{B}$ range from -15 eV to 20 eV, where the 3$d\rightarrow\ $2$p\ $XEPECS in Eq. (\ref{XEPECS}) was calculated for the $E_{B}$'s and the resulting spectra were collected according to Eq. (\ref{NXES}) for NXES.
The colored lines in Fig. \ref{Fig.2}(c) show the XEPECS spectra calculated at the binding energies indicated by the arrows with the same color in the 2$p$ XPS in Fig. \ref{Fig.2}(a). We clearly observe spectral narrowing of the XEPECS spectra from the NXES in Fig. \ref{Fig.2}(b).
This narrowing effect is a natural consequence of choosing a specific binding energy, i.e., a specific bunch of intermediate states, as discussed for rare-earth compounds in ref. 17.
Note that the intermediate states in the 3$d\rightarrow\ $2$p\ $NXES are the same as the final states in the 2$p$ XPS.
From Fig. \ref{Fig.2}(c), the main peak on the high-energy side of Fig. \ref{Fig.2}(b) is mainly contributed by the intermediate states higher in energy than those in the 2$p_{3/2}$ main line in Fig. \ref{Fig.2}(a), while the second peak on the low-energy side is mainly by the intermediate states around 2$p_{3/2}$ main line.
Here, we note that the XEPECS intensity becomes the strongest at $E_{B} = 2.64$ eV among the XEPECS spectra shown in Fig. \ref{Fig.2}(c). This can be explained as the combined effect between the 2$p$-3$d$ multiplet coupling and the phase matching of 3$d_{3/2, -1/2}$ state and 2$p_{1/2}$ state.
This narrowing effect is one of the advantages of the XEPECS over the NXES.\cite{STANAKA}

Another advantage of XEPECS is its ability to visualize spin and polarization entanglement through photoelectron-spin and emitted-photon-polarization resolved measurements.
Fig. \ref{Fig.2}(d) shows the result of the spin and polarization resolved 3$d\rightarrow\ $2$p\ $XEPECS calculated at $E_{B}$ = 2.64 eV in Fig. \ref{Fig.2}(c), where the four spectra are shown for the combination of the photoelectron spin (up/down) and the emitted photon polarization ($\lambda_{1}$/$\lambda_{2}$).
In Fig. \ref{Fig.2}(d), we observe characteristic intensity distributions of the four spectra.
In particular, for the main peak at $\omega$ = 7.49 eV, labeled "B", corresponding to the $bonding\ state$ between $\ket{d^{0}}$ and $\ket{d^{1}\underline{L}}$, we find only the spectra for up/$\lambda_{2}$ and down/$\lambda_{1}$ contribute to the peak intensity.
This suggests that the strong spin and polarization entanglement occurs at the bonding state in the spin and polarization resolved XEPECS at $E_{B}$ = 2.64 eV.
However, for the quantitative estimation of the degree of quantum entanglement, we need to calculate the density matrix in Eq. (\ref{rho}).
For this purpose, we calculate it with $E_{B}$ = 2.64 eV and $\omega$ = 7.49 eV, i.e., at the main peak B in Fig. \ref{Fig.2}(d).

The XEPECS structures in Fig. \ref{Fig.2}(d), the peak "B", satellite "NB" and "AB" correspond to the bonding, non-bonding and anti-bonding states, respectively.\cite{OKADA2}
Such a spectral structure is strongly affected by the polarization direction in the 3$d\rightarrow\ $2$p\ $RXES of $\rm TiO_{2}$.\cite{MATSUBARA2}
Actually, Matsubara et al. explained the strong polarization dependence found in the 3$d\rightarrow\ $2$p\ $RXES of $\rm TiO_{2}$ is due to the $a_{1g}$ symmetry of the bonding and the anti-bonding states between $\ket{d^{0}}$ and $\ket{d^{1}\underline{L}}$, i.e., the $a_{1g}$ symmetry of the final states strongly restricts the emitted-photon including polarization.\cite{MATSUBARA2}
Note that the final state of 3$d\rightarrow\ $2$p\ $XEPECS in $\rm Ti_{2}O_{3}$ is similar to the initial and final state of 3$d\rightarrow\ $2$p\ $RXES in $\rm TiO_{2}$. Thus, under the symmetry constraints as those in $\rm TiO_{2}$, strong spin and polarization entanglement is expected to occur at the bonding state because of the strong limitation on the spin and polarization degree of freedom.
In the subsequent discussion, we will further analyze this in terms of the density matrix.

\begin{figure}[H]
    \centering
    \includegraphics[scale=0.5]{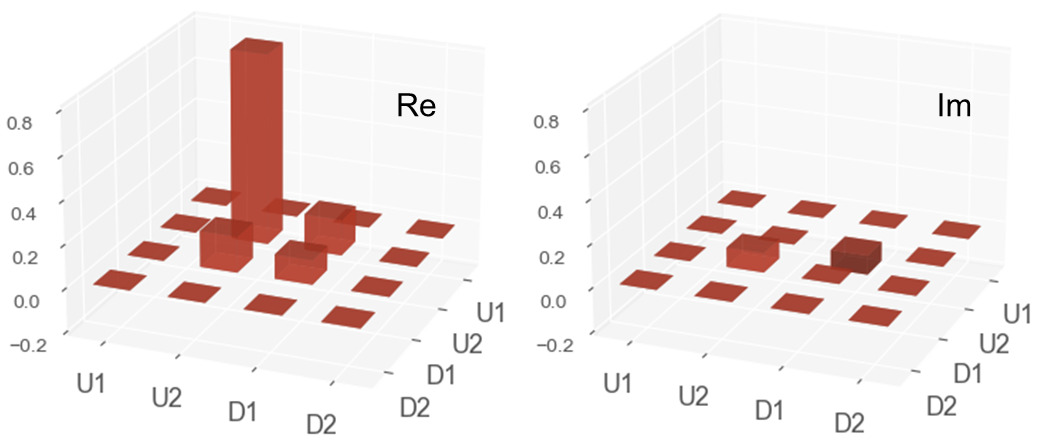}
    \caption{Real (Re) and imaginary (Im) parts of the density matrix for spin and polarization entanglement states calculated at the peak "B" in Fig.\ref{Fig.2}(d). The basis used here is the combined states between photoelectron spin (U for up and D for down) and emitted photon polarization (1 for $\lambda_{1}$ and 2 for $\lambda_{2}$).}
    \label{Fig.3}
\end{figure}

In Fig. \ref{Fig.3}, we show the calculated results of the real and imaginary parts of the density matrix at the peak "B" in Fig.\ref{Fig.2}(d) for the experimental setup in Fig. \ref{Fig.1}(b). 
Here, U and D denote the photoelectron up and down spin state, and 1 and 2 the linear polarization $\lambda_{1}$ and $\lambda_{2}$ of the emitted X-ray photon. We used the coupled states $\ket{\sigma\lambda}$ between the spin $\sigma$ and polarization $\lambda$ as the bases for calculating the density matrix.
The results clearly show the two off-diagonal elements, $\ket{U2}\bra{D1}$ and $\ket{D1}\bra{U2}$, together with the two diagonal elements, $\ket{U2}\bra{U2}$ and $\ket{D1}\bra{D1}$.
The existence of the finite off-diagonal element of $\ket{U2}\bra{D1}$ and $\ket{D1}\bra{U2}$ represent the coherence between the coupled states $\ket{U2}$ and $\ket{D1}$, i.e., the generated photoelectron and photon pair holds the spin and polarization entanglement.
In fact, assuming the spin and polarization maximally entangled state,
\begin{align}
\label{ideal func}
    \ket{\psi}=\frac{1}{\sqrt{2}}(\ket{U2}+e^{i\alpha}\ket{D1}),
\end{align}
we obtain fidelity $F\approx0.69$ at phase factor $\alpha\approx0.49$ using Eq. (\ref{F}). This $F$ value surpasses the classical limit of $0.5$, indicating the spin and polarization entanglement.
In addition, the tangle $T$, which represents the degree of quantum entanglement, becomes $T = 0.14$ in the present calculation.
This means that the quantum entanglement survives even in the presence of the charge-transfer effect as a disturbance in the realistic compounds, as discussed later in detail.

Now, we discuss the microscopic mechanism of the spin and polarization entanglement using the $\rm Ti^{3+}$ ionic model for simplicity. The Hamiltonian of the $\rm Ti^{3+}$ ion consists of the spin-orbit interaction and the crystal field. As mentioned before, the present study sets the initial state with the magnetic quantum number $M=-1/2$. Then, the ground state of the $\rm Ti^{3+}$ ion is expressed as
\begin{align}
    \label{g_ion} 
    \ket{g}_{\rm ion}=A_{1}\ket{d_{-1\uparrow}}-A_{2}\ket{d_{0\downarrow}},
\end{align}
where the coefficients $(A_{1}, A_{2})$ is given by $(\sqrt{3/5}, \sqrt{2/5}) \approx (0.77, 0.63)$ for $10Dq = 0.0$ eV and by $(0.99, 0.05)$ for $10Dq = 0.5$ eV.
As we discuss later, the coefficients $(A_{1}, A_{2})$ are important factors for the tangle describing the degree of the quantum entanglement. This means that the crystal field also affects the degree of quantum entanglement. 
However, the essential aspect of the spin and polarization entanglement is due to the spin-orbit interaction\cite{TOGAN,KOSAKA,TSURUMOTO,SEKIGUCHI,RTANAKA}.
Thus, in the following discussion, we restrict ourselves to consider only the spin-orbit interaction.

The core excitation to the intermediate state is described by
\begin{align}
\label{M_PEg}
    M_{PE}\ket{g}=\sum_{m,m',\sigma}(-1)^{m-m'}Y_{Dm}(\theta_{\rm e},\phi_{\rm e})e_{m'-m}\braket{D_{m}|C_{m-m'}^{1}|p_{m^{'}}}b_{m\sigma}^{\dag}p_{m'\sigma}\ket{g}_{\rm ion}.
\end{align}
Here, we used the formula of spherical harmonics expansion of the plain wave for the photoelectron shown in Fig. \ref{Fig.1}(b). Actually, we consider the most intense amplitude with the $D$-symmetry component of the photoelectron.
In Eq. (\ref{M_PEg}), the matrix element $\braket{D_{m}|C_{m-m'}^{1}|p_{m'}}$ is calculated following the Wigner-Eckart theorem and the coefficients $e_{m'-m}$ is given by 
\begin{align}
    \label{e_pm1}
    &e_{\pm1}\equiv\mp\frac{1}{\sqrt{2}}(\cos{\alpha}\cos{\theta}\pm i\sin{\alpha})e^{\pm i\phi},\\
    \label{e_0}
    &e_{0}\equiv-\cos{\alpha}\sin{\theta},
\end{align}
where $\theta$, $\phi$ corresponds to the incidence angles $(\theta_{\rm in}, \phi_{\rm in})$ of incident photon and the emission angles of emitted photon $(\theta_{\rm out}, \phi_{\rm out})$, and $\alpha$ describes the directions of the polarization vector of the incident photon and of the emitted photon, respectively.

Under the present setup in Fig. \ref{Fig.1}(b), Eq. (\ref{M_PEg}) is decomposed into 
\begin{align}
\label{M_PEg2}
    M_{PE}\ket{g}=(M_{PE}^{(1)}+M_{PE}^{(2)})\ket{g}_{\rm ion},
\end{align}
where 
\begin{align}
\label{M_PE1}
    &M_{PE}^{(1)}=-\sqrt{\frac{5}{32\pi}}i\bigg(\sqrt{3}\braket{D_{2}|C_{1}^{1}|p_{1}}b_{2\sigma}^{\dag}+\braket{D_{0}|C_{-1}^{1}|p_{1}}b_{0\sigma}^{\dag}\bigg)p_{1\sigma},\\
\label{M_PE2}
    &M_{PE}^{(2)}=-\sqrt{\frac{5}{32\pi}}i\bigg(\sqrt{3}\braket{D_{-2}|C_{-1}^{1}|p_{-1}}b_{-2\sigma}^{\dag}+\braket{D_{0}|C_{1}^{1}|p_{-1}}b_{0\sigma}^{\dag}\bigg)p_{-1\sigma}.
\end{align}
$M_{PE}^{(1)}$ and $M_{PE}^{(2)}$ mean the excitation of the Ti 2$p_{1\sigma}$ electron and the 2$p_{-1\sigma}$ electron, respectively. It should be noted that the Ti 2$p_{0\sigma}$ electron is not observed in the present setup in Fig. \ref{Fig.1}(b).
By inspecting Eqs. (\ref{M_PE1}) and (\ref{M_PE2}), we find that the total magnetic quantum number $M$ of the $\rm Ti^{3+}$ ion depends on the spin-up/down state of the photoelectron.
In the case of the up spin emission, $M$ takes the values $-2$ and $0$, and in the case of the down spin emission $-1$ and $1$.  
We list below the basis states for both cases:
\begin{itemize}
    \item[(1)] up spin PE
    \begin{itemize}
        \item[(A)] $M=-2$\\
        $\bigg\{
        \ket{\underline{p_{1\uparrow}}d_{-1\uparrow}}, \ket{\underline{p_{1\uparrow}}d_{0\downarrow}}, \ket{\underline{p_{0\uparrow}}d_{-1\downarrow}}, \ket{\underline{p_{0\uparrow}}d_{-2\uparrow}}, \ket{\underline{p_{-1\uparrow}}d_{-2\downarrow}}, \ket{\underline{p_{1\downarrow}}d_{-2\uparrow}}, \ket{\underline{p_{1\downarrow}}d_{-1\downarrow}},
        \ket{\underline{p_{0\downarrow}}d_{-2\downarrow}}
        \bigg\}$
        \item[(B)] $M=0$\\
        $\bigg\{
        \ket{\underline{p_{1\uparrow}}d_{1\uparrow}}, \ket{\underline{p_{1\uparrow}}d_{2\downarrow}}, \ket{\underline{p_{0\uparrow}}d_{0\uparrow}}, \ket{\underline{p_{0\uparrow}}d_{1\downarrow}}, \ket{\underline{p_{-1\uparrow}}d_{-1\uparrow}}, \ket{\underline{p_{-1\uparrow}}d_{0\downarrow}}, \ket{\underline{p_{1\downarrow}}d_{0\uparrow}}, \ket{\underline{p_{1\downarrow}}d_{1\downarrow}},\\
        \ket{\underline{p_{0\downarrow}}d_{-1\uparrow}}, \ket{\underline{p_{0\downarrow}}d_{0\downarrow}}, \ket{\underline{p_{-1\downarrow}}d_{-2\uparrow}}, \ket{\underline{p_{-1\downarrow}}d_{-1\downarrow}}
        \bigg\}$
    \end{itemize}
    \item[(2)]  down spin PE
    \begin{itemize}
        \item[(C)] $M=-1$\\
        $\bigg\{
        \ket{\underline{p_{1\uparrow}}d_{0\uparrow}}, \ket{\underline{p_{1\uparrow}}d_{1\downarrow}}, \ket{\underline{p_{0\uparrow}}d_{0\downarrow}}, \ket{\underline{p_{0\uparrow}}d_{-1\uparrow}}, \ket{\underline{p_{-1\uparrow}}d_{-1\downarrow}}, \ket{\underline{p_{-1\uparrow}}d_{-2\uparrow}}, \ket{\underline{p_{1\downarrow}}d_{0\downarrow}}, \ket{\underline{p_{1\downarrow}}d_{-1\uparrow}},\\
        \ket{\underline{p_{0\downarrow}}d_{-1\downarrow}}, \ket{\underline{p_{0\downarrow}}d_{-2\uparrow}}, \ket{\underline{p_{-1\downarrow}}d_{-2\downarrow}}
        \bigg\}$
        \item[(D)] $M=1$\\
        $\bigg\{
        \ket{\underline{p_{1\uparrow}}d_{2\uparrow}}, \ket{\underline{p_{0\uparrow}}d_{1\uparrow}}, \ket{\underline{p_{0\uparrow}}d_{2\downarrow}}, \ket{\underline{p_{-1\uparrow}}d_{0\uparrow}}, \ket{\underline{p_{-1\uparrow}}d_{1\downarrow}}, \ket{\underline{p_{1\downarrow}}d_{1\uparrow}}, \ket{\underline{p_{1\downarrow}}d_{2\downarrow}}, \ket{\underline{p_{0\downarrow}}d_{0\uparrow}},\\ \ket{\underline{p_{0\downarrow}}d_{1\downarrow}},
        \ket{\underline{p_{-1\downarrow}}d_{-1\uparrow}},
        \ket{\underline{p_{-1\downarrow}}d_{0\downarrow}}
        \bigg\}\hspace{0.2cm}$.
    \end{itemize}
\end{itemize}
Here, PE means the photoelectron, and $\ket{\underline{p_{m\sigma}}}$ and $\ket{d_{m\sigma}}$ denote the 2$p$ core-hole state and the 3$d$ state with orbital magnetic quantum number $m$ and spin magnetic quantum number $\sigma$.

The dipole transition matrix between the final state $\ket{f_{\sigma\lambda}}$ and intermediate basis states (A) to (D) is described by
\[
\left(
\begin{array}{c|c|c|c}
M_{ph}^{(M=-2 \text{ for } \uparrow \text{PE})} & 0 & \multicolumn{2}{c}{\text{\huge{0}}} \\ \cline{1-2}
0 & M_{ph}^{(M=0 \text{ for } \uparrow \text{PE})} & \multicolumn{2}{c}{} \\ \hline
\multicolumn{2}{c|}{} & M_{ph}^{(M=-1 \text{ for } \downarrow \text{PE})} & 0 \\ \cline{3-4}
\multicolumn{2}{c|}{\text{\huge{0}}} & 0 & M_{ph}^{(M=1 \text{ for } \downarrow \text{PE})} \\ 
\end{array}
\right),
\]
where the block matrices $M_{ph}^{(M=-2 \text{ for } \uparrow \text{PE})}$, $M_{ph}^{(M=0 \text{ for } \uparrow \text{PE})}$, $M_{ph}^{(M=-1 \text{ for } \downarrow \text{PE})}$ and $M_{ph}^{(M=1 \text{ for } \downarrow \text{PE})}$ are expressed using the coefficients of Eqs. (\ref{e_pm1}) and (\ref{e_0}) as follows:
\begin{align}
\label{M_{ph}^{(M=-2)}}
    &M_{ph}^{(M=-2 \text{ for } \uparrow \text{PE})}=
    \left(
    \begin{array}{ccc} \braket{f_{\uparrow\lambda_{1}}|M_{ph}|\underline{p_{1\uparrow}}d_{-1\uparrow}} & \dots & \braket{f_{\uparrow\lambda_{1}}|M_{ph}|\underline{p_{0\downarrow}}d_{-2\downarrow}}  \\ \braket{f_{\uparrow\lambda_{2}}|M_{ph}|\underline{p_{1\uparrow}}d_{-1\uparrow}} & \dots & \braket{f_{\uparrow\lambda_{2}}|M_{ph}|\underline{p_{0\downarrow}}d_{-2\downarrow}}
    \end{array}
    \right)\notag\\
    &=\left(
    \begin{array}{cccccccccc}
    0 & 0 & 0 & 0 & 0 & 0 & 0 & 0 \\
    0 & 0 & 0 & 0 & 0 & 0 & 0 & 0
    \end{array}
    \right),\\
\label{M_{ph}^{(M=0)}}
    &M_{ph}^{(M=0 \text{ for } \uparrow \text{PE})}=
    \left(
    \begin{array}{ccc} \braket{f_{\uparrow\lambda_{1}}|M_{ph}|\underline{p_{1\uparrow}}d_{1\uparrow}} & \dots & \braket{f_{\uparrow\lambda_{1}}|M_{ph}|\underline{p_{-1\downarrow}}d_{-1\downarrow}}\\ \braket{f_{\uparrow\lambda_{2}}|M_{ph}|\underline{p_{1\uparrow}}d_{1\uparrow}} & \dots & \braket{f_{\uparrow\lambda_{2}}|M_{ph}|\underline{p_{-1\downarrow}}d_{-1\downarrow}}
    \end{array}
    \right)\notag\\
    &=\sqrt{\frac{1}{5}}
    \left(
    \begin{array}{cccccccccccc}
    e_{0}^{(\lambda_{1})} & 0 & \frac{2}{\sqrt{3}}e_{0}^{(\lambda_{1})} & 0 & e_{0}^{(\lambda_{1})} & 0 & 0 & e_{0}^{(\lambda_{1})} & 0 & \frac{2}{\sqrt{3}}e_{0}^{(\lambda_{1})} & 0 & e_{0}^{(\lambda_{1})} \\
    e_{0}^{(\lambda_{2})} & 0 & \frac{2}{\sqrt{3}}e_{0}^{(\lambda_{2})} & 0 & e_{0}^{(\lambda_{2})} & 0 & 0 & e_{0}^{(\lambda_{2})} & 0 & \frac{2}{\sqrt{3}}e_{0}^{(\lambda_{2})}& 0 & e_{0}^{(\lambda_{2})}
    \end{array}
    \right),\\
\label{M_{ph}^{(M=-1)}}
    &M_{ph}^{(M=-1 \text{ for } \downarrow \text{PE})}=
    \left(
    \begin{array}{ccc} \braket{f_{\downarrow\lambda_{1}}|M_{ph}|\underline{p_{1\uparrow}}d_{0\uparrow}} & \dots & \braket{f_{\downarrow\lambda_{1}}|M_{ph}|\underline{p_{-1\downarrow}}d_{-2\downarrow}}\\ \braket{f_{\downarrow\lambda_{2}}|M_{ph}|\underline{p_{1\uparrow}}d_{0\uparrow}} & \dots & \braket{f_{\downarrow\lambda_{2}}|M_{ph}|\underline{p_{-1\downarrow}}d_{-2\downarrow}}
    \end{array}
    \right)\notag\\
    &=\sqrt{\frac{1}{5}}
    \left(
    \begin{array}{ccccccccccc}
    \sqrt{\frac{1}{3}}e_{-1}^{(\lambda_{1})} & 0 & 0 & e_{-1}^{(\lambda_{1})} & 0 & \sqrt{2}e_{-1}^{(\lambda_{1})} & \sqrt{\frac{1}{3}}e_{-1}^{(\lambda_{1})} & 0 & e_{-1}^{(\lambda_{1})} & 0 & \sqrt{2}e_{-1}^{(\lambda_{1})}\\
    \sqrt{\frac{1}{3}}e_{-1}^{(\lambda_{2})} & 0 & 0 & e_{-1}^{(\lambda_{2})} & 0 & \sqrt{2}e_{-1}^{(\lambda_{2})} & \sqrt{\frac{1}{3}}e_{-1}^{(\lambda_{2})} & 0 & e_{-1}^{(\lambda_{2})} & 0 & \sqrt{2}e_{-1}^{(\lambda_{2})}
    \end{array}
    \right),\\ \rm and\notag\\
\label{M_{ph}^{(M=1)}}
    &M_{ph}^{(M=1 \text{ for } \downarrow \text{PE})}=
    \left(
    \begin{array}{ccc} \braket{f_{\downarrow\lambda_{1}}|M_{ph}|\underline{p_{1\uparrow}}d_{2\uparrow}} & \dots & \braket{f_{\downarrow\lambda_{1}}|M_{ph}|\underline{p_{-1\downarrow}}d_{0\downarrow}}\\ \braket{f_{\downarrow\lambda_{2}}|M_{ph}|\underline{p_{1\uparrow}}d_{2\uparrow}} & \dots & \braket{f_{\downarrow\lambda_{2}}|M_{ph}|\underline{p_{-1\downarrow}}d_{0\downarrow}}
    \end{array}
    \right)\notag\\
    &=\sqrt{\frac{1}{5}}
    \left(
    \begin{array}{ccccccccccc}
    \sqrt{2}e_{1}^{(\lambda_{1})} & e_{1}^{(\lambda_{1})} & 0 & \sqrt{\frac{1}{3}}e_{1}^{(\lambda_{1})} & 0 & 0 & \sqrt{2}e_{1}^{(\lambda_{1})} & 0 & e_{1}^{(\lambda_{1})} & 0 & \sqrt{\frac{1}{3}}e_{1}^{(\lambda_{1})}\\
    \sqrt{2}e_{1}^{(\lambda_{2})} & e_{1}^{(\lambda_{2})} & 0 & \sqrt{\frac{1}{3}}e_{1}^{(\lambda_{2})} & 0 & 0 & \sqrt{2}e_{1}^{(\lambda_{2})} & 0 & e_{1}^{(\lambda_{2})} & 0 & \sqrt{\frac{1}{3}}e_{1}^{(\lambda_{2})}
    \end{array}
    \right).
\end{align}
Here, we should note that all the elements of $M_{ph}^{(M=-2 \text{ for } \uparrow \text{PE})}$ are zeros under the geometry in Fig. \ref{Fig.1}(b), because of the dipole selection rule. 
To explain this, we take the element $\braket{f_{\uparrow\lambda_{1}}|M_{ph}|\underline{p_{1\uparrow}}d_{-1\uparrow}}$ of $M_{ph}^{(M=-2 \text{ for } \uparrow \text{PE})}$, as an example. 
In this case, the $d_{-1\uparrow}$ electron cannot decay into the $p_{1\uparrow}$ hole because $\Delta m$ exceeds the limit of the dipole transition, and thus the element becomes 0.
As shown in Fig. \ref{Fig.1}(b), we set the emission angle of the emitted photon as ($\theta_{\rm out}, \phi_{\rm out}) =(90^{\circ}, 45^{\circ}$) and the polarization $\lambda_{1}$ and $\lambda_{2}$ as $\alpha=90^{\circ}$ and $\alpha=180^{\circ}$, respectively.
Under the present setup, the first-row elements of $M_{ph}^{(M=0 \text{ for } \uparrow \text{PE})}$ in Eq. (\ref{M_{ph}^{(M=0)}}) becomes zero, as well as the second-row elements of $M_{ph}^{(M=-1 \text{ for } \downarrow \text{PE})}$ and of $M_{ph}^{(M=1 \text{ for } \downarrow \text{PE})}$ in Eqs. (\ref{M_{ph}^{(M=-1)}}) and (\ref{M_{ph}^{(M=1)}}), because $e_{m'-m}^{(\lambda)}$ vanishes.
As a result, the elements corresponding to only the pair of the up-spin PE and the polarization $\lambda_{2}$ and the pair of the down-spin and $\lambda_{1}$ remain as finite values.
Thus, the off-diagonal elements $\ket{U2}\bra{D1}$ and $\ket{D1}\bra{U2}$ in Fig. \ref{Fig.3} occur due to the structure of the dipole transition matrix under the setup in Fig. \ref{Fig.1}(b), in addition to the spin-orbit interaction.

The structure of the dipole transition matrix also explains the strong dependence on the setup. For example, the degree of the spin and polarization entanglement changes as a function of the emission angle $\theta_{\rm out}$ and vanishes at $\theta_{\rm out} = 0^{\circ}$ and $180^{\circ}$. This reflects that the second-row elements of $M_{ph}^{(M=0 \text{ for } \uparrow \text{PE})}$ become 0 and only some elements of $M_{ph}^{(M=-1 \text{ for } \downarrow \text{PE})}$ and $M_{ph}^{(M=1 \text{ for } \downarrow \text{PE})}$ survives.
In fact, the tangle ($T$) takes the value 0 in these cases.

In the real part of Fig. \ref{Fig.3}, we find an asymmetry between the diagonal elements $\ket{U2}\bra{U2}$ and $\ket{D1}\bra{D1}$. This asymmetry can be qualitatively understood by considering the two-step XEPECS process.
First, the incident photon excites a 2$p$ core electron, creating a 2$p$ core hole. The photoelectron's spin is approximately aligned with the 2$p$ core-hole's spin, even in the presence of the 2$p$ spin-orbit interaction which can flip the 2$p$ core-hole spin.
Actually, numerical calculation tells us that the 2$p$ core-hole spin occupancy is $( \braket{p_{\uparrow}}, \braket{p_{\downarrow}}) = (0.62, 0.38)$ when the photoelectron spin is up in the experimental setup in Fig. \ref{Fig.1}(b). 
In the case of the down spin photoelectron, on the other hand the occupancy becomes $(0.35,0.65)$.
Second, radiative relaxation occurs when the 3$d$ electron spin matches the 2$p$ core-hole spin because of the dipole selection rule. In Eq. (\ref{g_ion}), the coefficients $A_1$ and $A_2$ for $\ket{d_{-1\uparrow}}$ and $\ket{d_{0\downarrow}}$ determine the values of the diagonal elements $\ket{U2}\bra{U2}$ and $\ket{D1}\bra{D1}$. Since $A_1 > A_2$, $\ket{U2}\bra{U2}$ becomes larger than $\ket{D1}\bra{D1}$, leading to the observed asymmetry.
As discussed below, this asymmetry contributes to the reduction of spin and polarization entanglement.

\begin{figure}[H]
    \centering
    \includegraphics[scale=0.38]{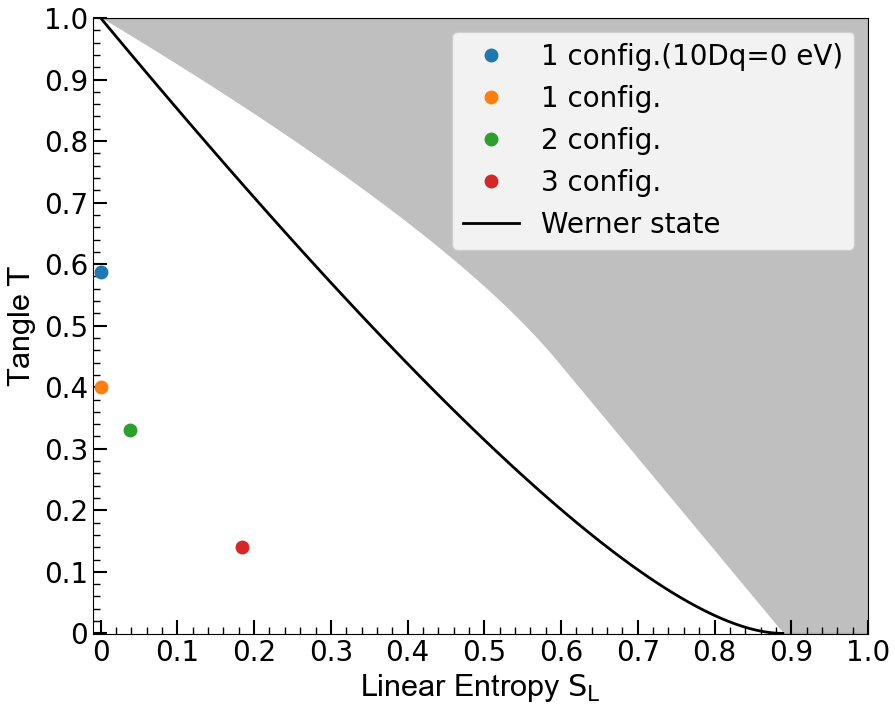}
    \caption{Relationship between tangle $T$ and linear entropy $S_{L}$ using the $\rm TiO_{6}$-type cluster model with various numbers of electronic configuration (config.) included in the simulation: one config. ($\ket{d^{1}}$) with crystal field splitting energy $10Dq = 0$ eV, one config. ($\ket{d^{1}}$), two configs. ($\ket{d^{1}}, \ket{d^{2}\underline{L}}$) and three configs ($\ket{d^{1}}, \ket{d^{2}\underline{L}}, \ket{d^{3}\underline{L}^{2}}$) with $10Dq = 0.5$ eV. The solid line indicates the Werner states, and the grey area corresponds to physically impossible states.\cite{MUNRO2,PETERS}}
    \label{Fig.4}
\end{figure}

Finally, to examine the charge-transfer effect on the spin and polarization entanglement in the 3$d \rightarrow$ 2$p$ XEPECS process, we calculate the tangle $T$ and the linear entropy $S_L$ using the density matrix in Eq. (\ref{rho}).
The tangle $T$ quantifies the degree of quantum entanglement, while the linear entropy $S_L$ measures the mixedness of the density matrix. They both range from 0 to 1: a higher $T$ indicates stronger entanglement, whereas $S_{L}$ changes from 0 to 1 when the system changes from a pure state to a maximally mixed state.
Fig. \ref{Fig.4} shows the tangle and linear entropy for the $\rm TiO_6$ cluster calculated with different numbers of basis configurations. In the case of the single configuration calculation, i.e., $\rm Ti^{3+}$ ion case, $T$ is at its highest, and $S_L = 0$, indicating a pure state. When the number of configurations increases to two and three, $T$ decreases while $S_L$ increases, reflecting a larger mixing effect of states due to the hybridization between Ti 3$d$ and O 2$p$.
In addition to hybridization, the crystal field also influences the spin and polarization entanglement. As shown in Fig. \ref{Fig.4}, when the crystal field splitting energy $10Dq$ is set to 0 eV, $T$ for a single configuration is higher than that for the same configuration with $10Dq = 0.5$ eV. This effect is linked to the asymmetry of the diagonal elements in the density matrix. Specifically, the crystal field increases the ratio of $A_1$ to $A_2$ in Eq. (\ref{g_ion}), leading to this asymmetry.
As a result, the enhanced asymmetry reduces the tangle.
In addition to the analysis of charge-transfer effects via the increase in the number of electronic configurations, we also investigate the dependence of the tangle $T$  and the linear entropy $S_{L}$ on the charge transfer energy $\Delta$ for three configurations($\ket{d^{1}}, \ket{d^{2}\underline{L}}, \ket{d^{3}\underline{L}^{2}}$). The results are summarized in Table \ref{table_dlt}. 
As shown in Table \ref{table_dlt}, $T$ increases with increasing $\Delta$ while $S_{L}$ decreases. 
The increase in the charge-transfer energy $\Delta$ leads to an increase in the weight $\alpha_{0}$ of the $\ket{d^{1}}$ configuration, while the weights $\alpha_{1}$ and $\alpha_{2}$ of the $\ket{d^{2}\underline{L}}$ and $\ket{d^{3}\underline{L}^{2}}$ configurations decrease, as defined in Eq. (\ref{g_clu}).
In other words, increasing $\Delta$ drives the system closer to the Ti$^{3+}$ ion case with single configuration ($\ket{d^{1}}$) and suppresses charge-transfer effects.
As discussed in the configuration-number dependence, the entanglement becomes stronger(i.e., the tangle $T$ increases) and the linear entropy $S_L$ decreases as the system approaches the single-configuration.
Therefore, the analysis based on the increase in charge-transfer energy $\Delta$ leads to the same conclusion as the configuration-number dependence.

\begin{table}[H]
\centering
\begin{tabular}{ccccccccccc}
\hline\hline
$\Delta$ (eV) & $\ $ & 5.5 & $\ $ & 6.0 & $\ $ & 6.5 & $\ $ & 7.0 & $\ $ & 7.5 \\
\hline
$T$ & $\ $ & 0.10 & $\ $ & 0.12 & $\ $ & 0.14 & $\ $ & 0.15 & $\ $ & 0.17 \\
$S_{L}$ & $\ $ & 0.22 & $\ $ & 0.21 & $\ $ & 0.18 & $\ $ & 0.17 & $\ $ & 0.15 \\
\hline\hline
\end{tabular}
\caption{Charge transfer energy $\Delta$-dependence of the tangle $T$ and the linear entropy $S_{L}$ for the $\rm TiO_{6}$-type cluster model with three configurations ($\ket{d^{1}}, \ket{d^{2}\underline{L}}, \ket{d^{3}\underline{L}^{2}}$).}
\label{table_dlt}
\end{table}

While the present study focus on the $M = -1/2$ subspace at $\phi_{\rm out} = 45^{\circ}$, we also calculated the tangle values for the other $M$ subspaces ($M = -3/2, -1/2, 1/2, 3/2$) at $\phi_{\rm out} = 0^{\circ}$ and $45^{\circ}$, as shown in Table \ref{table_mj}. These results indicate that finite entanglement can also emerge in the other $M$ subspaces depending on the emission angle $\phi_{\rm out}$.
We also note that the entangled spin–polarization pair depends on both $M$ and the emission geometry: for example, up spin/$\lambda_1$ and down spin/$\lambda_2$ pairs appear for $M = -3/2$ and $1/2$, whereas up/$\lambda_2$ and down/$\lambda_1$ pairs appear for $M = -1/2$ and $3/2$.
A comprehensive analysis of the $M$- and angle-dependence of the entanglement will be an important subject of future work.

\begin{table}[H]
\centering
\begin{tabular}{c|cc}
\hline\hline
$M$ & $T\ (\phi_{\rm out} = 0^\circ)$ & $T\ (\phi_{\rm out} = 45^\circ)$ \\
\hline
$-3/2$ & 0.06 & 0.02 \\
$-1/2$ & 0.13 & 0.14 \\
$1/2$  & 0.13 & 0.40 \\
$3/2$  & 0.06 & 0.05 \\
\hline\hline
\end{tabular}
\caption{Tangle $T$ for each initial state labeled by magnetic quantum number $M$ at emission angles $\phi_{\rm out} = 0^\circ$ and $45^\circ$.}
\label{table_mj}
\end{table}

\section{Conclusion}
In this study, we have theoretically investigated the quantum entanglement between the spin of the photoelectron and the polarization of the emitted X-ray photon in the 3$d \rightarrow$ 2$p$ XEPECS process for the target compound $\rm Ti_{2}O_{3}$. Using the $\rm TiO_{6}$-type cluster model with $O_{h}$ symmetry, we considered the full-multiplet structure of the $\rm Ti$ ion and the charge-transfer effect between the $\rm Ti$ 3$d$ and $\rm O$ 2$p$ orbitals. Our results provide a deeper understanding of how quantum entanglement manifests in the XEPECS process under the realistic electronic structures influenced by both intra-atomic interactions and ligand effects.

We demonstrated that, for the experimental setup depicted in Fig. \ref{Fig.1}(b), the degree of entanglement for the up spin/$\lambda_{2}$ and the down spin/$\lambda_{1}$ pairs takes a value of 0.13. Furthermore, we analytically showed that the entanglement vanishes ($T = 0$) for the emission angles of the X-ray photon at $\theta_{\rm out} = 0^{\circ}$ and $180^{\circ}$ in the case of the isolated $\rm Ti^{3+}$ ion, which is similar to the case of our simpler treatment \cite{RTANAKA}.
The present study, incorporating a more realistic cluster model, reinforces the generality of the angular dependence of spin and polarization entanglement in the XEPECS process. These findings suggest that the detection geometry strongly affects the degree of quantum entanglement in X-ray spectroscopic measurements.

In addition to the emission-angle dependence, we found that the charge-transfer effect between the $\rm Ti$ 3$d$ and $\rm O$ 2$p$ orbitals significantly influences the degree of entanglement. Actually, the $\rm Ti$ 3$d$ - $\rm O$ 2$p$ hybridization reduces the degree of entanglement as the electronic configuration changes from a single-configuration picture to a multi-configuration state. This result also indicates that the spin and polarization entanglement state is disturbed by the hybridization effect, resulting in a reduction in the purity of the entangled state. 
In addition, we found that the increased asymmetry of the diagonal elements of the real part of the density matrix because the crystal field further suppresses the entanglement. 
These results highlight the strong dependence of quantum entanglement on both charge-transfer and crystal-field effects, which are intrinsic to real materials beyond simple atomic models.

This study belongs to the field of X-ray quantum optics, as it considers the entire system, including the photon degrees of freedom. The theoretical framework extends conventional X-ray spectroscopy by incorporating quantum entanglement as a key observable. The spin and polarization entanglement properties in the XEPECS process provide a basis for developing novel spectroscopic techniques that exploit quantum correlations in the X-ray regime. Systematic investigations of various material systems will lead to a deeper understanding of the fundamental mechanisms governing X-ray-induced quantum entanglement and contribute to the advancement of methods for probing electronic states in condensed matter physics.

\begin{acknowledgment}
%\acknowledgment

We would like to thank Professor Satoshi Tanaka for valuable discussions. This work was supported by JST SPRING, Grant Number JPMJSP2139.

\end{acknowledgment}

\end{document}